\documentclass[aps,prb,twocolumn,superscriptaddress,showpacs]{revtex4}

\usepackage{graphicx}
\usepackage{dcolumn}
\usepackage{bm}

\begin{document}

\title{Magnetically originated phonon-glass-like behavior in
Tb$_2$Ti$_2$O$_7$ single crystal}

\author{Q. J. Li}
\affiliation{Hefei National Laboratory for Physical Sciences at
Microscale, University of Science and Technology of China, Hefei,
Anhui 230026, People's Republic of China}

\author{Z. Y. Zhao}
\affiliation{Hefei National Laboratory for Physical Sciences at
Microscale, University of Science and Technology of China, Hefei,
Anhui 230026, People's Republic of China}

\author{C. Fan}
\affiliation{Hefei National Laboratory for Physical Sciences at
Microscale, University of Science and Technology of China, Hefei,
Anhui 230026, People's Republic of China}

\author{F. B. Zhang}
\affiliation{Hefei National Laboratory for Physical Sciences at
Microscale, University of Science and Technology of China, Hefei,
Anhui 230026, People's Republic of China}

\author{H. D. Zhou}
\affiliation{Department of Physics and Astronomy, University of
Tennessee, Knoxville, Tennessee 37996-1200, USA}
\affiliation{National High Magnetic Field Laboratory, Florida
State University, Tallahassee, Florida 32306-4005, USA}

\author{X. Zhao}
\email{xiazhao@ustc.edu.cn} \affiliation{School of Physical
Sciences, University of Science and Technology of China, Hefei,
Anhui 230026, People's Republic of China}

\author{X. F. Sun}
\email{xfsun@ustc.edu.cn} \affiliation{Hefei National Laboratory
for Physical Sciences at Microscale, University of Science and
Technology of China, Hefei, Anhui 230026, People's Republic of
China}

\date{\today}

\begin{abstract}

We report a study on the thermal conductivity ($\kappa$) of
Tb$_2$Ti$_2$O$_7$ single crystals at low temperatures. It is found
that in zero field this material has an extremely low phonon
thermal conductivity in a broad temperature range. The mean free
path of phonons is even smaller than that of amorphous materials
and is 3--4 orders of magnitude smaller than the sample size at
0.3 K. The strong spin fluctuation of the spin-liquid state is
discussed to be the reason of the strong phonon scattering. The
magnetic-field dependence of $\kappa$ and comparison with
Y$_2$Ti$_2$O$_7$ and TbYTi$_2$O$_7$ confirm the magnetic origin of
this phonon-glass-like behavior.

\end{abstract}


\maketitle

\section{INTRODUCTION}

Ultra-low thermal conductivity ($\kappa$) of crystal lattice or
phonons is a long-term investigated topic.\cite{Snyder, Hsu,
Chiritescu, Costescu} It is directly related to the applications
of thermoelectric materials and thermal barrier materials.
Introducing structural imperfections into the samples, such as
impurities and defects, is always a useful method to increase the
phonon scattering and decrease the phonon thermal
conductivity.\cite{Berman} In an extreme case, the amorphous
solids are known to have very weak heat transport of
phonons.\cite{Berman} Because of the lack of the long-range
periodicity of the atom positions in amorphous materials, the
phonon mean free path is very short even at very low
temperatures.\cite{Berman} However, in many cases, it is not a
suitable way to obtain small phonon thermal conductivity by simply
using amorphous materials. For example, the highly efficient
thermoelectric materials should have simultaneous low thermal
conductivity and high electric conductivity.\cite{Snyder, Hsu}
This requirement cannot be met in those samples with significant
crystal imperfections or in amorphous solids. In other words, to
obtain very small phonon thermal conductivity in bulk single
crystals is still a challenging task.

There may be some chances in magnetic materials since some recent
works have revealed that the magnetic fluctuations could yield
strong scattering on phonons. The magnetic excitations can work as
either heat carriers or phonon scatterers.\cite{Berman} In those
materials exhibiting long-range magnetic order, the strong
magnetic fluctuations near the critical region of phase transition
can scatter phonons rather strongly.\cite{Berman, Wang_HMO,
Ke_BMO, Chen_MCCL, Zhao_GFO, Zhao_NCO} A recent example is the
hexagonal manganite HoMnO$_3$.\cite{Wang_HMO} The phonon thermal
conductivity was found to be suppressed by two orders of magnitude
at the regime of antiferromagnetic (AF) transition of Ho$^{3+}$
ions. However, such scattering is not effective at lower
temperatures when the long-range order is well
established.\cite{Ke_BMO, Chen_MCCL, Zhao_GFO, Zhao_NCO} In
addition, the phonon thermal conductivities of these materials are
still much larger than those of the amorphous materials. Even
stronger magnetic scattering effect is therefore to be expected in
those materials with strong spin fluctuations and no long-range
order. This is what the spin-liquid materials may exhibit. Another
requirement for strong phonon scattering in magnetically
disordered state is that the coupling between spin and lattice
should be strong.

Geometrical spin frustration is very promising to result in a
three-dimensional spin liquid.\cite{Balents} A well studied
material is the rare-earth pyrochlore
Tb$_2$Ti$_2$O$_7$.\cite{Gardner1, Gardner2, Cao1, Takatsu, Petti,
Ruff1, Ruff2, Molvavian1, Molvavian2, Lhotel, Fennell, Legel,
Yin1} Although the ground state of this materials is still a
controversial question, most of the previous studies evidenced
that it is a cooperative paramagnet down to milli-kelvin
temperatures, much lower than the energy scale of the AF
interactions. The single-ion ground state is a degenerate doublet
and the Tb$^{3+}$ spin has an Ising-like anisotropy with the easy
axis along the local [111] axis.\cite{Cao1} In addition, the
peculiar crystal field effect yields a pronounced magnetoelastic
effect, which was verified by X-ray diffraction and demonstrated
that the coupling between spin and lattice is very
strong.\cite{Ruff1} As a result, an anomalously strong structural
fluctuation has been observed in zero field.\cite{Ruff2}
Therefore, Tb$_2$Ti$_2$O$_7$ seems to meet the requirements for
exhibiting strong spin-phonon scattering. In this work, the
thermal conductivity of high-quality Tb$_2$Ti$_2$O$_7$ single
crystals is measured down to 0.3 K. It is found that in a broad
temperature regime the thermal conductivity is extremely small and
comparable to that of the amorphous materials. In particular, the
mean free path of phonons is found to be as low as 20 nm at 4 K
and 3--4 orders of magnitude smaller than the sample size at
subkelvin temperatures, and is even shorter than that of amorphous
materials.\cite{Berman} This result shows an exceptional case of
very low phonon thermal conductivity in an insulating single
crystal. The strong spin fluctuation in the spin-liquid-like state
is the main mechanism of the phonon scattering. Both the
magnetic-field-induced increase of $\kappa$ and the comparison of
Tb$_2$Ti$_2$O$_7$ with Y$_2$Ti$_2$O$_7$ and TbYTi$_2$O$_7$ also
confirm the magnetic origin of this phonon-glass-like behavior.

\section{EXPERIMENTS}

High-quality Tb$_2$Ti$_2$O$_7$, Y$_2$Ti$_2$O$_7$, and
TbYTi$_2$O$_7$ single crystals were grown by the floating-zone
technique.\cite{Li_Crystal} These crystals could be grown well
under different oxygen pressures and growth rates.
Tb$_2$Ti$_2$O$_7$ and TbYTi$_2$O$_7$ crystals were grown in 0.4
MPa pure oxygen with a growth rate of 2.5 mm/h and in 0.25 MPa
pure oxygen with a rate of 3 mm/h, respectively. Y$_2$Ti$_2$O$_7$
crystal was grown in mixed oxygen and argon with a ratio of 4:1
and at a growth rate of 4 mm/h. The thermal conductivities were
measured using a conventional steady-state technique and two
different processes: (i) using a ``one heater, two thermometers"
technique in a $^3$He refrigerator and a 14 T magnet at
temperature regime of 0.3 -- 20 K; (ii) using a Chromel-Constantan
thermocouple in a pulse-tube refrigerator for zero-field data
above 5 K.\cite{Sun_DTN, Wang_HMO, Ke_BMO, Chen_MCCL, Zhao_GFO,
Zhao_NCO} Note that a careful precalibration of resistor (RuO$_2$)
sensors is indispensable for the precise thermal conductivity
measurements in high magnetic fields and at low temperatures. The
specific heat was measured by the relaxation method in the
temperature range from 0.4 to 30 K using a commercial physical
property measurement system (PPMS, Quantum Design).

\section{RESULTS AND DISCUSSION}

\begin{figure}
\includegraphics[clip,width=7cm]{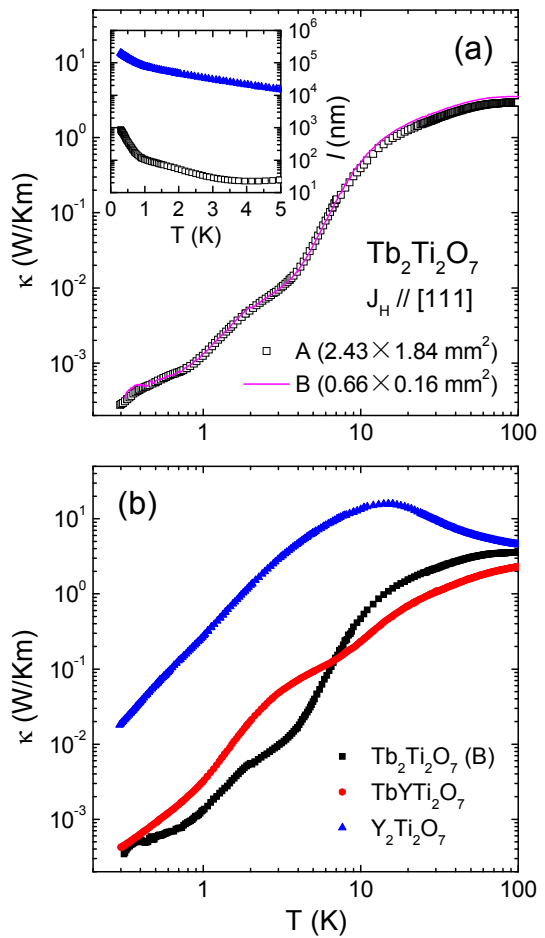}
\caption{(color online) (a) Zero-field thermal conductivities of
two Tb$_2$Ti$_2$O$_7$ single crystals with very different sizes.
The width and thickness of these two samples are displayed. (b)
Comparison of the zero-field thermal conductivity among
Tb$_2$Ti$_2$O$_7$, Y$_2$Ti$_2$O$_7$, and TbYTi$_2$O$_7$ single
crystals having similar sample widths. Inset to panel (a) shows
the phonon mean free paths of Tb$_2$Ti$_2$O$_7$ (sample $A$) and
Y$_2$Ti$_2$O$_7$.}
\end{figure}

Figure 1(a) shows the zero-field thermal conductivities of two
Tb$_2$Ti$_2$O$_7$ samples with very different dimensions
(2.91$\times$0.66$\times$0.16 mm$^3$ and
3.31$\times$2.43$\times$1.84 mm$^3$). The heat currents were
applied along the length direction, which was along the [111] axis
of crystal lattice. There are several remarkable features in these
two curves that point to a phonon-glass-like behavior of these
single-crystal samples. First, the magnitudes of $\kappa$ are very
small and there is no any signature of the so-called phonon peaks
at low temperatures. It is well known that the $\kappa(T)$ of
insulators usually exhibits a pronounced peak at low temperatures
(10--20 K),\cite{Berman} which is a characteristic of phonon heat
transport. The absence of phonon peak in a high-quality single
crystal is actually quite unusual and indicates a strong phonon
scattering effect. In addition, the overall temperature dependence
of $\kappa$ is very similar to what the amorphous solids exhibit,
in which the phonons behave as a glassy state. Second, the mean
free path of phonons is found to be extremely short in
Tb$_2$Ti$_2$O$_7$ crystals. The phononic thermal conductivity can
be expressed by the kinetic formula $\kappa_{ph} =
\frac{1}{3}Cv_pl$,\cite{Berman} where $C = \beta T^3$ is the
phonon specific heat at low temperatures, $v_p$ is the average
velocity and $l$ is the mean free path of phonons. Using the
$\beta$ value obtained from specific-heat
measurements,\cite{Li_Crystal, Specific_Heat} the phonon velocity
can be calculated and then the mean free path is obtained from the
$\kappa$.\cite{Zhao_GFO, Zhao_NCO} The inset to Fig. 1(a) shows
the calculated $l$ at low temperatures of the big sample, which is
only $\sim$ 20 nm at 4 K. Even when lowering temperature to 0.3 K,
$l$ is shorter than 1000 nm and is 3--4 orders of magnitude
smaller than the geometrical size of sample. Note that the mean
free path is so short that it is even smaller than the typical
magnitudes in the amorphous materials.\cite{Berman} In contrast,
the microscopic phonon scatterings in usual single crystals, such
as the phonon-phonon scattering and scattering by various crystal
defects, would be quenched at very low temperatures and the mean
free path could reach the sample size. This is known as the phonon
boundary scattering limit.\cite{Berman} It means that at
temperatures as low as 0.3 K the phonon scattering in
Tb$_2$Ti$_2$O$_7$ crystals is still very strong. Note that this
cannot be caused by the crystal defects since the temperature is
low enough. Third, the comparison of two samples with very
different sizes also confirmed the absence of phonon boundary
scattering. In usual crystals, when the phonon mean free path is
long enough to be comparable to the sample size, the thermal
conductivity is proportional to the sample size.\cite{Berman}
However, the two sets of data in Fig. 1(a) indicate that the
$\kappa$ values are essentially independent on the size, which
also evidences that the mean free path of phonons must be
significantly shorter than the sample size.

It is necessary to discuss the role of magnetic excitations in the
heat transport of Tb$_2$Ti$_2$O$_7$. Although the nature of the
ground state is not fully understood, it is most likely a
spin-liquid-like disordered state, in which the magnetic
excitations are short-range correlated spin fluctuations and were
evidenced by the low-energy spectra of neutron
measurements.\cite{Ruff3} From the extremely low thermal
conductivity of Tb$_2$Ti$_2$O$_7$, it can be easily concluded that
the spin fluctuations strongly scatter phonons rather than act as
heat carriers.

Figure 1(b) compares the thermal conductivity of Tb$_2$Ti$_2$O$_7$
with Y$_2$Ti$_2$O$_7$ and TbYTi$_2$O$_7$ single crystals.
Y$_2$Ti$_2$O$_7$ has the same crystal structure with
Tb$_2$Ti$_2$O$_7$ but is nonmagnetic. It is clearly seen that the
thermal conductivity of Y$_2$Ti$_2$O$_7$ (having size of
4.51$\times$0.63$\times$0.15 mm$^3$) behaves exactly like the
usual insulting crystals, with a phonon peak at about 15 K. For a
more quantitative comparison, the phonon mean free path of
Y$_2$Ti$_2$O$_7$ crystal was also calculated and shown in the
inset to Fig. 1(a).\cite{Specific_Heat_2} At low temperatures, it
is 2--3 orders of magnitude larger than that of Tb$_2$Ti$_2$O$_7$
and at 0.3 K it reaches a large value ($\sim$ 2$\times$10$^5$ nm),
close to the averaged sample width, suggesting that the boundary
scattering limit is almost established. Apparently, the small
difference of lattice parameters of Y$_2$Ti$_2$O$_7$ and
Tb$_2$Ti$_2$O$_7$ cannot lead to so large difference of phonon
transport. The main difference between these two materials is
their magnetisms. Therefore, it is naturally concluded that the
magnetic scattering, due to the strong magnetic fluctuations in
the quantum-spin-liquid state of Tb$_2$Ti$_2$O$_7$, is the main
mechanism for the extremely strong phonon scattering. The heat
transport of another material, TbYTi$_2$O$_7$, seems to support
this point. TbYTi$_2$O$_7$ is stoichiometrically a mixture of
Tb$_2$Ti$_2$O$_7$ and Y$_2$Ti$_2$O$_7$, which would be expected to
exhibit a weaker heat transport than two parent
compounds.\cite{Berman} The high-temperature ($>$ 10 K) data
indeed show such a result. However, at very low temperatures the
$\kappa$ of TbYTi$_2$O$_7$ is clearly larger than that of
Tb$_2$Ti$_2$O$_7$. Since the dilution of Tb$^{3+}$ ions by the
nonmagnetic Y$^{3+}$ ions would suppress the spin correlations,
the magnetic scattering on phonons is reasonably weakened.

\begin{figure}
\includegraphics[clip,width=8.5cm]{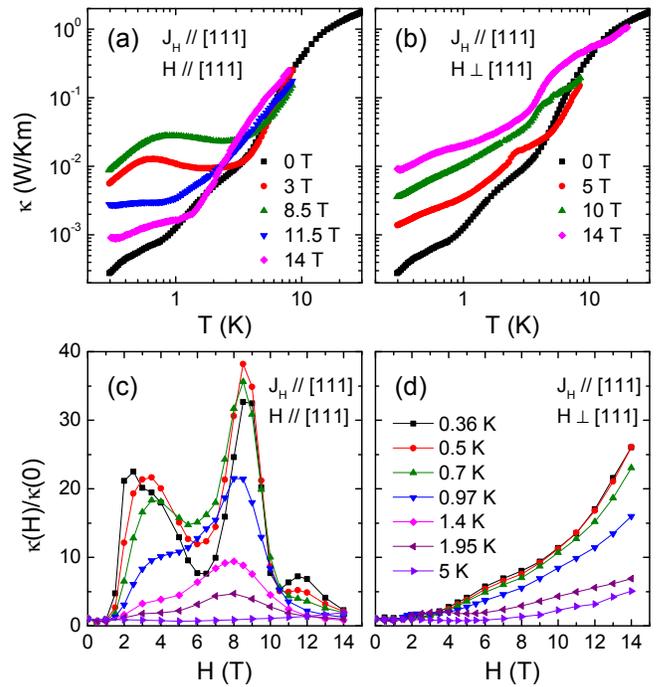}
\caption{(color online) Temperature and field dependencies of
$\kappa$ of Tb$_2$Ti$_2$O$_7$ single crystal for magnetic field
along (a,c) and perpendicular to (b,d) the [111] axis,
respectively. All the data were taken on one sample ($A$). The
magnetic field perpendicular to the [111] direction is actually
along another special direction [1$\bar{1}$0].}
\end{figure}

If the strong phonon scattering of Tb$_2$Ti$_2$O$_7$ single
crystals is magnetically originated, the magnetic field is
expected to be able to strongly affect the heat transport, which
is confirmed by the field dependence measurements of $\kappa$.
Figure 2 shows how the magnetic field, parallel to either the
[111] direction or the vertical one, changes the thermal
conductivity. A striking finding is that at very low temperatures
the $\kappa$ can be enhanced by 30--40 times with increasing
field. To our knowledge, there has been no report of such a strong
magnetothermal conductivity effect in any other magnetic
materials. In principle, it is understandable that the magnetic
field can suppress the spin fluctuations and therefore weaken the
phonon scattering. In this sense, a 30--40 times increase of
$\kappa$ in fields is actually far below the expectation.
Considering that the mean free path of phonons is 3--4 orders of
magnitude smaller than the geometrical size of the sample, one may
expect more than 1000 times enhancement of $\kappa$ if the
magnetic field is strong enough to smear out all the magnetic
scattering on phonons.

It is notable that the field dependencies are quite different for
two field directions. For $H \perp$ [111], the very-low-$T$
$\kappa$ is nearly field independent for $H <$ 2 T and then
increases monotonically with field. It seems that the magnetic
field along this direction just simply suppresses the spin
fluctuations. The effect of magnetic field along [110] direction
had been well studied,\cite{Ruff3, Cao2} in which case the
Tb$^{3+}$ spins were thought to lying on two sets of chains along
the [110] and [1$\bar{1}$0] directions. The spins on chains
parallel to the field direction (so-called $\alpha$ chains) align
along the local [111] axis with a component parallel to the
magnetic field. Whereas the spins on chains perpendicular to the
field direction (the $\beta$ chains) favor an AF order at high
fields. This field-induced order had already been revealed by the
neutron scattering.\cite{Ruff3, Cao2} It was found that the
field-induced elastic neutron scattering intensity increased
gradually above 2 T and at subkelvin temperatures, which is
qualitatively consistent with the increase of $\kappa(H)$ above 2
T. At higher temperatures ($>$ 3 K), the field-induced magnetic
order was hardly to be observed;\cite{Ruff3, Cao2}
correspondingly, the $\kappa$ shows much weaker enhancement with
increasing field.

For $H \parallel$ [111], both the field and temperature
dependencies show much more complicated behaviors, compared to the
case of $H \perp$ [111]. The 0.36-K $\kappa(H)$ curve shows a
three-peak (at 2.5, 8.5, and 11.5 T) or three-dip (at 0.5, 6, and
10.5 T) feature between 0 and 14 T (see Fig. 3(d) also). In
principle, the field-induced enhancement of $\kappa$ in this case
is likely to have the same mechanism as the case of $H \perp$
[111], i.e., some field-induced magnetic order. However, it is
hard to image so complicated magnetic transitions if each anomaly
on the $\kappa(H)$ curves corresponds to a magnetic transition.
There actually has been few experimental studies on the possible
field-induced changes of magnetic properties for $H
\parallel$ [111]. An early neutron scattering experiment had
found that the magnetic field along the [111] axis could also
induce an AF order.\cite{Yasui} The elastic neutron scattering
intensity was found to increase with field up to 2--3 T and become
nearly saturated, which seems to have some correspondence with the
sharp increase of $\kappa$ at 1--2 T. However, either the nature
or the spin structure of this ordered phase were not
resolved.\cite{Yasui} Magnetic susceptibility data taken at
several tens of milli-kelvin temperatures also indicated a
shoulder-like anomaly at about 1.2 T,\cite{Legel, Yin1} which is
likely related to a field-induced order. More recent experiments
indicated that this anomaly exists and shows a weak temperature
dependence at temperature scale of several hundred
milli-kelvins.\cite{Yin2} However, such susceptibility
measurements did not reveal any other clear anomaly for fields up
to 14 T.\cite{Yin2} Based on these existing experimental results,
it is not likely that there are some multiple magnetic transitions
in high fields along the [111] axis. A recent susceptibility study
has revealed that Tb$_2$Ti$_2$O$_7$ is likely to exhibit a
quantum-spin-ice ground state and the magnetic field along the
[111] axis can drive transitions very similar to those in the
spin-ice material Dy$_2$Ti$_2$O$_7$.\cite{Yin1} However, these
transitions between the ``quantum-spin-ice", the
``quantum-kagom\'e-ice", and the ``three-in-one-out" states occur
in rather low fields ($<$ 1 T) and at very low temperatures ($<$
150 mK).\cite{Yin1} Therefore, these low-field transitions are
apparently irrelevant to the $\kappa(H)$ behaviors in Fig. 2(c).
There should be some other reasons for the complicated $\kappa(H)$
behavior besides the field-induced orders or transitions.

\begin{figure}
\includegraphics[clip,width=8.5cm]{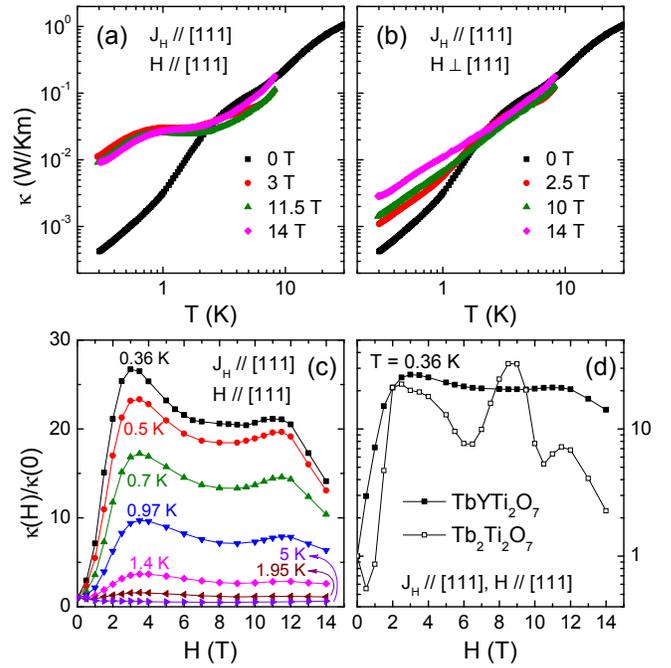}
\caption{(color online) (a,b) Temperature dependencies of $\kappa$
of TbYTi$_2$O$_7$ single crystal for magnetic field along and
perpendicular to the [111] axis, respectively. (c) Magnetic field
dependencies of $\kappa$ of TbYTi$_2$O$_7$ single crystal for $H
\parallel$ [111]. (d) Comparison of the $\kappa(H)$ data at 0.36 K
between Tb$_2$Ti$_2$O$_7$ and TbYTi$_2$O$_7$.}
\end{figure}

Similar field-dependence measurements have been done on the
TbYTi$_2$O$_7$ single crystal, as shown in Fig. 3. The effects of
magnetic field in this sample are also very strong, but weaker
than those in Tb$_2$Ti$_2$O$_7$. It is understandable if the low
thermal conductivity of Tb$_2$Ti$_2$O$_7$ is due to the strong
magnetic scattering on phonons and TbYTi$_2$O$_7$ has weaker
magnetic fluctuations. Nevertheless, there are several remarkable
phenomena in the transport properties of TbYTi$_2$O$_7$. First, as
shown in Fig. 3(b), the high-field induced enhancement of $\kappa$
for $H \perp$ [111] is several times weaker in TbYTi$_2$O$_7$ than
that in Tb$_2$Ti$_2$O$_7$. Since half of the Tb$^{3+}$ ions are
replaced by Y$^{3+}$, the exchange interactions between magnetic
ions are weakened and it is probably difficult to form the
field-induced magnetic order of the $\beta$ chains. In this sense,
the stronger high-field enhancement of $\kappa$ in
Tb$_2$Ti$_2$O$_7$ has some relationship to the spin orders, which
can result in weaker magnetic scattering than the isolated
magnetic ions in TbYTi$_2$O$_7$. Second, the $\kappa(H)$ isotherms
for $H \parallel$ [111] show comparably large magnetothermal
conductivity but rather different field dependencies, compared
with those of Tb$_2$Ti$_2$O$_7$. In particular, each low-$T$
$\kappa(H)$ curve displays two rather broad peaks at about 3 and
11.5 T and a shallow valley at 8--9 T. However, the two peak
positions are nearly the same as two of the Tb$_2$Ti$_2$O$_7$
peaks, as shown in Fig. 3(d). This indicates that the $\kappa(H)$
behaviors in $H \parallel$ [111] may have close relationship to
the single-ion properties, like the crystal-field excitations.

\begin{figure}
\includegraphics[clip,width=6cm]{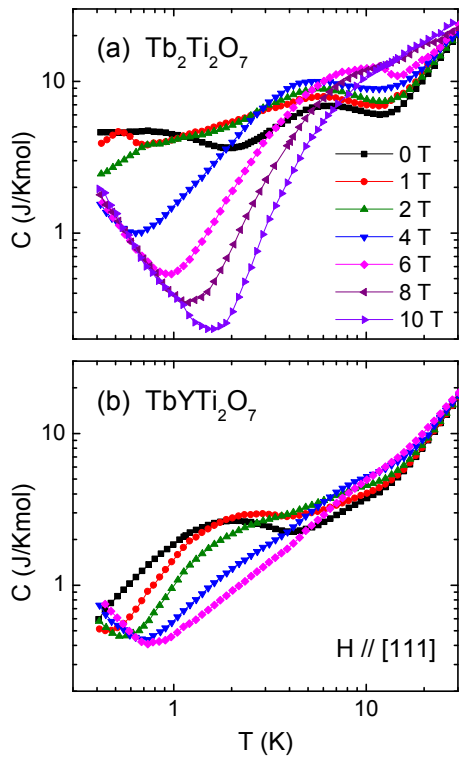}
\caption{(color online) Low-temperature specific heat of
Tb$_2$Ti$_2$O$_7$ and TbYTi$_2$O$_7$ single crystals in magnetic
field along the [111] axis.}
\end{figure}

Figure 4 shows the specific heat of Tb$_2$Ti$_2$O$_7$ and
TbYTi$_2$O$_7$ single crystals at low temperatures and in the
[111] fields. Our zero-field specific heat data of
Tb$_2$Ti$_2$O$_7$ are essentially consistent with an earlier
work.\cite{Gingras} Two broad peaks or humps at about 0.7 and 6 K
could be related to the ground-state and the first-excitation
doublets of Tb$^{3+}$ crystal-field levels, broadened by the
exchange interactions.\cite{Gingras, Hamaguchi, Takatsu} In
magnetic fields along the [111] axis, the two anomaly peaks react
sensitively to the field. In high fields, only the high-$T$ peak
remains and shifts to higher temperatures, while the low-$T$ one
disappears at fields above 2 T. Note that the lowest-$T$ upturns
of specific heat in high fields is likely some nuclear spin
contribution.\cite{Gingras, Hamaguchi} In TbYTi$_2$O$_7$, the
low-$T$ anomaly moved significantly to $\sim$ 2 K and the high-$T$
one moves slightly to $\sim$ 8 K and becomes much weaker in the
magnitude. Nevertheless, the specific-heat data do not show any
obvious signature for the possible magnetic order and the
field-induced changes are mainly related to the crystal-field
excitations.

Let us discuss about the possibility of the crystal-field effects
on the exotic field dependencies of $\kappa$ in the [111] field.
Since the low-field transition to some AF order state can
significantly weaken the phonon scattering, a monotonic increase
of $\kappa$ with increasing field is expected, similar to the case
of $H \perp$ [111]. Thus, the high-field behavior of $\kappa$ can
occur if there are resonant phonon scattering by the crystal-field
excitations\cite{Berman, Sun_PLCO, Sun_GBCO, Li_NGSO} at those
particular fields like 6 and 10.5 T, and probably another one
higher than 14 T. In this regard, the crystal-field levels of
Tb$_2$Ti$_2$O$_7$ were rather clearly determined at least at 4 K.
The ground state and the first excitation are non-Kramers doublets
with $\sim$ 18 K separation, and higher levels are
singlets.\cite{Gingras, Mirebeau, Lummen, Malkin} Considering the
Zeeman splitting of the doublets, one can figure out two possible
phonon resonant scatterings at 0.36 K. The first one is that the
splitting of the ground-state doublet can produce a resonant
scattering at about 0.2 T, when the energy splitting is about
3.8$k_BT$.\cite{Berman, Sun_PLCO, Sun_GBCO, Li_NGSO} The second
one is expected when the lower branch of the first-excitation
level becomes close enough to the ground-state level (with
interval of 3.8$k_BT$), which occurs at about 16 T. Note that due
to the strong anisotropy of the Tb$^{3+}$ spins, the Land\'e
factor is about 10--14 for the [111] direction and zero for the
direction perpendicular to the [111] axis. \cite{Malkin}
Therefore, the resonant scattering by the crystal-field
excitations can work only for fields along [111]. It is reasonable
that only with $H \parallel$ [111] the $\kappa(H)$ and high-field
$\kappa(T)$ shows the dip-like features, which could be due to the
resonant scattering.\cite{Berman, Sun_PLCO, Sun_GBCO, Li_NGSO}
However, the above estimated resonant scattering fields are
significantly different from the dip fields in $\kappa(H)$.

Another possible reason for the anisotropic field dependence of
$\kappa$ is related to the magnetostriction, which was
experimentally found to be strongly anisotropic
also.\cite{Aleksandrov, Klekovkina} The existing data indicated
that the longitudinal and transverse magnetostrictions only showed
smooth functions of field up to 6 T at 4 K. However, since the
magnetoelastic coupling is quickly enhanced with lowering
temperatures,\cite{Ruff2} some drastic field dependencies of
magnetostriction can be expected at subkelvin temperatures. When
the remarkable changes of lattice parameters and elastic constants
are induced by the magnetic field, lowering crystal symmetry and
structural phase transitions could occur in strong
field.\cite{Klekovkina} Such effects on phonon heat transport was
recently observed in another rare-earth pyrochlore
Dy$_2$Ti$_2$O$_7$.\cite{Fan_DTO, Kolland} Nevertheless,
very-low-$T$ measurements of the magnetostriction would be useful
for clarifying the mechanism of the magnetothermal conductivity
effect in Tb$_2$Ti$_2$O$_7$.

\section{SUMMARY}

An extremely low thermal conductivity was observed in
Tb$_2$Ti$_2$O$_7$ single crystals. The short mean free path of
phonons points to a phonon-glass-like behavior of this material.
Both the field dependencies of $\kappa$ and the comparison with
Y$_2$Ti$_2$O$_7$ and TbYTi$_2$O$_7$ indicate that the phonons are
strongly scattered by the spin fluctuations, whereas the
field-induced changes of magnetism are remained to be
investigated.

\begin{acknowledgments}

We thank L. Yin for helpful discussions. This work was supported
by the National Natural Science Foundation of China, the National
Basic Research Program of China (Grant Nos. 2009CB929502 and
2011CBA00111), and the Fundamental Research Funds for the Central
Universities (Program No. WK2340000035).

\end{acknowledgments}


\begin{thebibliography}{}
\bibitem{Specific_Heat}
Since the specific heat of Tb$_2$Ti$_2$O$_7$ has a significant
contribution from magnetic excitations, as shown in Fig. 4(a), it
is difficult to directly get the phonon specific-heat coeffcient
$\beta$. An alternative way is to refer the data of
Lu$_2$Ti$_2$O$_7$,\cite{Li_Crystal} in which the Lu$^{3+}$ ion is
nonmagnetic and has similar ion radius and mass to Tb$^{3+}$. The
low-$T$ specific heat data of Lu$_2$Ti$_2$O$_7$ can be well fitted
by using the formula $C = \beta T^3 + \beta_5 T^5+\beta_7 T^7$,
with $\beta = 1.55 \times 10^{-3}$ J/K$^4$mol.
\bibitem{Specific_Heat_2}
Our specific heat data of Y$_2$Ti$_2$O$_7$ show a simple lattice
contribution at low temperatures.\cite{Li_Crystal} Fitting the
data to the formula $C = \beta T^3 + \beta_5 T^5+\beta_7 T^7$
gives $\beta = 2.36 \times 10^{-4}$ J/K$^4$mol. This value is
considerably smaller than that of Lu$_2$Ti$_2$O$_7$. It seems to
be also smaller than an earlier result of
Y$_2$Ti$_2$O$_7$,\cite{Gingras} which gave $\beta = 1.25 \times
10^{-3}$ J/K$^4$mol if one used the same fitting formula. Even
though this $\beta$ value corresponds to a phonon mean free path 3
times smaller than that in Fig. 1(a), our conclusion is
essentially unchanged, considering that the difference between
Tb$_2$Ti$_2$O$_7$ and Y$_2$Ti$_2$O$_7$ is several orders of
magnitude.
\end{thebibliography}
\end{document}